\documentclass[runningheads,a4paper]{llncs}
\usepackage{epsfig} 
\usepackage{subfigure}
\usepackage{multirow}
\usepackage{natbib}

\usepackage{hyperref}

\usepackage{amssymb}
\usepackage{url}
\urldef{\mailsa}\path|{alfred.hofmann, ursula.barth, ingrid.haas, frank.holzwarth,|
\urldef{\mailsb}\path|anna.kramer, leonie.kunz, christine.reiss, nicole.sator,|
\urldef{\mailsc}\path|erika.siebert-cole, peter.strasser, lncs}@springer.com|
\newcommand{\keywords}[1]{\par\addvspace\baselineskip
\noindent\keywordname\enspace\ignorespaces#1}

\usepackage{pifont}
\usepackage{rotating}
\usepackage{enumerate}
\usepackage{float}
\usepackage{amsmath,graphicx}
\usepackage{array,booktabs}
\usepackage{verbatim}
\newcommand{\like}{\log P}

\DeclareMathOperator*{\argmax}{\mathrm{argmax}}

\newcommand{\Poi}{\text{Poi}}
\newcommand{\psihat}{\hat{\psi}}
\newcommand{\psiout}{\psi^{\mathrm{out}}}
\newcommand{\thetahat}{\hat{\theta}}
\newcommand{\thetaout}{\theta^{\mathrm{out}}}
\newcommand{\thetain}{\theta^{\mathrm{in}}}
\newcommand{\alphaout}{\alpha^{\mathrm{out}}}
\newcommand{\alphain}{\alpha^{\mathrm{in}}}
\newcommand{\betaout}{\beta^{\mathrm{out}}}
\newcommand{\betain}{\beta^{\mathrm{in}}}
\newcommand{\lambdaout}{\lambda^{\mathrm{out}}}
\newcommand{\lambdain}{\lambda^{\mathrm{in}}}
\newcommand{\thetahatout}{\hat{\theta}^{\mathrm{out}}}
\newcommand{\thetahatin}{\hat{\theta}^{\mathrm{in}}}
\newcommand{\kappaout}{\kappa^{\mathrm{out}}}
\newcommand{\kappain}{\kappa^{\mathrm{in}}}
\newcommand{\dout}{d^{\mathrm{out}}}
\newcommand{\din}{d^{\mathrm{in}}}

\newcommand{\xmin}{x_{\mathrm{min}}}
\newcommand{\xmax}{x_{\mathrm{max}}}
\newcommand{\MLEs}{\mbox{MLEs}}
\newcommand{\MLE}{\mbox{MLE}}
\newcommand{\MCMC}{\mbox{MCMC}}
\newcommand{\NMI}{\mbox{NMI}}
\newcommand{\SBM}{\mbox{SBM}}
\newcommand{\DG}{\mbox{DG}}
\newcommand{\DC}{\mbox{DC}}
\newcommand{\DDC}{\mbox{DDC}}
\newcommand{\ODC}{\mbox{ODC}}
\newcommand{\DGDC}{\mbox{DG-DC}}
\newcommand{\DGODC}{\mbox{DG-ODC}}
\newcommand{\DGDDC}{\mbox{DG-DDC}}
\newcommand{\NH}{\mbox{NH}}

\newcommand{\e}{\mathrm{e}}

\newcommand{\dtheta}{{\mathrm d}\theta}
\newcommand{\dpsi}{{\mathrm d}\psi}

\renewcommand{\mid}{\,|\,}

\newcommand{\thetamin}{\theta_{\min}}
\newcommand{\thetaminr}{\theta_{\min,r}}

\begin{document}

\mainmatter  

\title{Oriented and Degree-generated Block Models:\\
Generating and Inferring Communities\\
with Inhomogeneous Degree Distributions}

\titlerunning{Degree Correction in Stochastic Block Models}

%
%
\author{Yaojia Zhu\inst{1}%
\thanks{This work was supported by the McDonnell Foundation.}%
\and Xiaoran Yan\inst{1} \and Cristopher Moore\inst{2,1}}
\authorrunning{Yaojia Zhu, Xiaoran Yan, Cristopher Moore}

\institute{University of New Mexico\\ Albuquerque NM 87131 USA
\and Santa Fe Institute\\ 1399 Hyde Park Road, Santa Fe NM 87501, USA}

%
%

\maketitle

\begin{abstract}
The stochastic block model is a powerful tool for inferring community structure from network topology. However, it predicts a Poisson degree distribution within each community, while most real-world networks have a heavy-tailed degree distribution.  The degree-corrected block model can accommodate arbitrary degree distributions within communities. But since it takes the vertex degrees as parameters rather than generating them, it cannot use them to help it classify the vertices, and its natural generalization to directed graphs cannot even use the orientations of the edges. In this paper, we present variants of the block model with the best of both worlds: they can use vertex degrees and edge orientations in the classification process, while tolerating heavy-tailed degree distributions within communities. We show that for some networks, including synthetic networks and networks of word adjacencies in English text, these new block models achieve a higher accuracy than either standard or degree-corrected
block models.
\end{abstract}

\section{Introduction}

In many real-world networks, vertices can be divided into \emph{communities} or \emph{modules} based on their connections. Social networks can be forged by interactions in daily activities like karate training~\cite{zachary}. The blogosphere contains groups of linked blogs with similar political views~\cite{ADAMIC:2005:ID38}.  Words can be tagged as different parts of speech based on their adjacencies in large texts~\cite{newman-leicht}.  Communities range from assortative clumps, where vertices preferentially attach to others of the same type, to \emph{functional} communities of vertices that connect to the rest of the network in similar ways, such as groups of predators in a food web that feed on similar prey~\cite{Allesina,moore_2011}.  Understanding this variety of community structures, and their relationships to functional roles of vertices and edges,
is crucial to understanding network data.

The \emph{stochastic block model} (\SBM)~\cite{FIENBERG:1981:ID001,HOLLAND:1983:ID73,SNIJDERS:1997:ID168,AIROLDI:2008:ID72} is a popular and highly flexible generative model for community detection. It partitions the vertices into communities or \emph{blocks}, where vertices belonging to the same block are \emph{stochastically equivalent}~\cite{WASSERMAN:1987:ID172} in the sense that the probabilities of a connection with all other vertices are the same for all vertices in the same block.  This definition of community is quite flexible,
letting block models capture many types of community structure, including assortative, disassortative, and satellite communities and mixtures of them~\cite{NEWMAN:2002:ID79,NEWMAN:2003:ID171,MORUP:2009:ID40,moore_2011,decelle_2011,decelle_2011_2}.

The \SBM\ assumes that each edge is generated independently conditioned on the block memberships.  Each entry $A_{uv}$ of the adjacency matrix is then Bernoulli-distributed, where the probability that $A_{uv}=1$ depends solely on the block memberships $g_u, g_v$ of its endpoints. Since
every pair of vertices in a given pair of blocks are connected with the same probability, for large $n$ the degree distribution within each block is Poisson. As a consequence, vertices with very different degrees are unlikely to be in the same block. This leads to problems when modeling real networks, which often have heavy-tailed degree distributions within each community.  For instance, both liberal and conservative political blogs range from high-degree ``leaders'' to low-degree ``followers''~\cite{ADAMIC:2005:ID38}.

Recently, Karrer and Newman~\cite{KARRER:2011:ID106} developed the \emph{degree-corrected} (\DC) block model for undirected networks. They add a parameter for each vertex, which controls its expected degree.
By setting these parameters equal to the observed degrees, the DC can accommodate arbitrary degree distributions within communities.  This removes the model's tendency to separate high-degree and low-degree vertices into different communities. Similar models were considered by M{\o}rup and Hansen~\cite{MORUP:2009:ID40} and Reichardt, Alamino, and Saad~\cite{Reichardt2011}.

On the other hand, the degree-corrected model cannot use the vertex degrees to help it classify the vertices, precisely because it takes the degrees as parameters rather than as data that need to be explained.  For this reason, \DC\ may actually fail to recognize communities that differ significantly in their degree distributions.  Thus we have two extremes: the \SBM\ separates vertices by degree even when it shouldn't, and DC fails to do so even when it should.

For directed graphs, the natural generalization of \DC, the \emph{directed degree-corrected} (\DDC) block model, has two parameters for each vertex: the expected in-degree and out-degree.  But this model cannot even take advantage of edge orientations.  For instance, in English adjectives usually precede nouns but rarely vice versa.  Thus the ratio of each vertex's in- and out-degree is strongly indicative of its block membership, and leveraging this part of the data is very helpful in the classification process.

In this paper, we propose two new types of block model, which combine the strengths of the degree-corrected and uncorrected block models.  The \emph{oriented degree-corrected} (\ODC) block model is able to utilize the edge orientations for community detection by only correcting the total degrees instead of the in- and out-degrees separately.  We show that for networks with strongly asymmetric behavior between communities, including synthetic networks and networks of word adjacencies in English text, \ODC\ achieves a higher accuracy than either the original stochastic block model or the degree-corrected block model.

We also propose the \emph{degree-generated} (\DG) block model, which treats the expected degree of each vertex as generated from prior distributions in each community, such as power laws whose exponents and cutoffs vary from one community to another.  By including the probability of these degrees in the likelihood of a given block assignment, the model captures the interaction between the degree distribution and the community structure. \DG\ automatically strikes a balance between allowing vertices of different degrees to coexist in the same community on the one hand, and using vertex degrees to separate vertices into communities on the other.

Our experiments show that \DG\ works especially well in networks where communities have highly inhomogeneous degree distributions, but where the degree distributions differ enough between communities so that we can use vertex degrees to help us classify the vertices.  Both the standard and degree-corrected block models classify nodes solely on the basis of the relative density of connections between communities, with different notions of ``density.''  \DG\ block models let us leverage degree information as well.
In some cases, \DG\ has a further advantage in faster convergence as it reshapes the landscape of the parameter space, providing the inference algorithm a shortcut to the correct community structure.

These new variants of the block model give us the best of both worlds.  They can tolerate heavy-tailed degree distributions within communities, but can also use degrees and edge orientations to help classify the vertices. In addition to their performance on these networks, our models illustrate a valuable point about generative models and statistical inference: when inferring the structure of a network, you can only use the information that you try to generate.

\section{The models}

In this section, we review the degree-corrected block model of~\cite{KARRER:2011:ID106}, and present our variations on it, namely oriented and degree-generated block models.

\subsection{Background: degree-corrected block models}
\label{sec:dc}

Throughout, we use $N$ and $M$ to denote the number of vertices and edges, and $K$ to denote the number of blocks.  The problem of determining the number of blocks is a subtle model selection problem, which we do not address here.

In the original stochastic block model, the entries $A_{uv}$ of the adjacency matrix are independent and Bernoulli-distributed, with $P(A_{uv}=1)=p_{g_u,g_v}$.  Here $g_u$ is the block to which $u$ belongs, where $p$ is a $K \times K$ matrix.  Karrer and Newman~\cite{KARRER:2011:ID106} consider random multigraphs where the $A_{uv}$ are independent and Poisson-distributed,
\[
A_{uv} \sim \Poi(\theta_u \theta_v \omega_{g_u,g_v}) \, .
\]
Here $\omega$ replaces $p$, and $\theta_u$ is an overall propensity for $u$ to connect to other vertices. Note that since the $A_{uv}$ are independent, the degrees $d_u$ will vary somewhat around their expectations; however, the resulting model is much simpler to analyze than one that controls the degree of each vertex exactly.

Ignoring self-loops, the likelihood with which this degree-corrected (\DC) block model generates an undirected multigraph $G$ is then
\begin{align}
\label{eq:lh_dc} P(G\mid \theta,\omega, g)
= \prod_{u < v}\frac{\left(\theta_u\theta_v\omega_{g_ug_v}\right)^{A_{uv}}}{A_{uv}!}\exp\left(-\theta_u\theta_v\omega_{g_ug_v}\right)
 \, .
\end{align}
To remove the obvious symmetry where we multiply the $\theta$'s by a constant $C$ and divide $\omega$ by $C^2$, we can impose a normalization constraint $\sum_{u:g_u=r}\theta_u=\kappa_r$ for each block $r$, where $\kappa_r=\sum_{u:g_u=r} d_u$ is the total degree of the vertices in block $r$. Under these constraints, the maximum likelihood estimates (\MLEs) for the $\theta$ parameters are $\hat{\theta}_u=d_u$. For each pair of blocks $r, s$, the \MLE\ for $\omega_{rs}$ is then
\[
\hat{\omega}_{rs} = \frac{m_{rs}}{\kappa_r \kappa_s} \, ,
\]
where $m_{rs}$ is the number of edges connecting block $r$ to block $s$ (and edges within blocks are counted twice).  Substituting these \MLEs\ for $\theta$ and $\omega$ then gives the log-likelihood
%
\begin{align}
\label{eq:llh_dc} \like(G \mid g)
= \frac{1}{2}\sum_{r,s=1}^K m_{rs} \log \frac{m_{rs}}{\kappa_r\kappa_s} \, .
\end{align}

\subsection{Directed and oriented degree-corrected models}
\label{sec:ddc-odc}

The natural extension of the degree-corrected model to directed networks, which we call the directed degree-corrected block model (\DDC), has two parameters $\thetaout_u,\thetain_u$ for each vertex.  The number of directed edges from $u$ to $v$ is again Poisson-distributed,
\[
A_{uv} \sim \Poi(\thetaout_u \thetain_v \omega_{g_u,g_v}) \, .
\]
We impose the constraints $\sum_{u:g_u=r} \thetaout_u=\kappaout_r$ and $\sum_{u:g_u=r} \thetain_u=\kappain_r$ for each block $r$,
where $\kappaout_r = \sum_{u:g_u=r} \dout_u$ and $\kappain_r = \sum_{u:g_u=r} \din_u$ denote the total out- and in-degree of block $r$.
As before, let $m_{rs}$ denote the number of directed edges from block $r$ to block $s$.  Then the likelihood is
\begin{align}
\label{eq:lh_doi}
P(G\mid\theta,\omega,g)
&= \prod_{uv}\frac{\left(\thetaout_u\,\thetain_v\,\omega_{g_ug_v}\right)^{A_{uv}}}{A_{uv}!}\exp(-\thetaout_u\,\thetain_v\,\omega_{g_ug_v}) \nonumber\\
&= \frac{\prod_u(\thetaout_u)^{\dout_u}(\thetain_u)^{\din_u}\prod_{rs}\omega_{rs}^{m_{rs}}\exp(-\kappaout_r\kappain_s\omega_{rs})}{\prod_{uv}A_{uv}!}\, ,
\end{align}
Ignoring constants, we get the log-likelihood as follows
\begin{align}
\label{eq:llh1_doi}
\log P(G\mid\theta,\omega,g)
&= \sum_u(\dout_u\log\thetaout_u +\din_u\log\thetain_u)\nonumber\\
&+ \sum_{rs}(m_{rs}\log\omega_{rs}-\kappaout_r\kappain_s\omega_{rs}) \, .
\end{align}
The \MLEs\ for the parameters (see Appendix~\ref{app:ddc}) are
\begin{align}
\label{eq:mle_doi}
\thetahatout_u=\dout_u \, , \quad
\thetahatin_u=\din_u \, , \quad
\hat{\omega}_{rs}=\frac{m_{rs}}{\kappaout_r\kappain_s} \, .
\end{align}
Substituting these \MLEs\ gives
\begin{align}
\label{eq:llh2_doi}
\log P(G\mid g)=\sum_{r,s=1}^K m_{rs} \log\frac{m_{rs}}{\kappaout_r \kappain_s} \, .
\end{align}

In the \DDC, the in- and out-degrees of each vertex are completely specified by the $\theta$ parameters, at least in expectation.  Thus the \DDC\ lets vertices with arbitrary in- and out-degrees to fit comfortably together in the same block.  On the other hand, since the degrees are given as parameters, rather than as data that the model must generate and explain, the \DDC\ cannot use them to infer community structure.  Indeed, it cannot even take advantage of the orientations of the edges, and as we will see below it performs quite poorly on networks with strongly asymmetric community structure.

To deal with this, we present a partially degree-corrected block model capable of taking advantage of edge orientations, which we call the \emph{oriented degree-corrected} (\ODC) block model. Following the maxim that we can only use the information that we try to generate, we correct only for the total degrees of the vertices, and generate the edges' orientations.

Let $\bar{G}$ denote the undirected version of a directed graph $G$, i.e., the multigraph resulting from erasing the arrows for each edge.  Its adjacency matrix is $\bar{A}_{uv}=A_{uv}+A_{vu}$, so (for instance) $\bar{G}$ has two edges between $u$ and $v$ if $G$ had one pointing in each direction.  The \ODC\ can be thought of as generating $\bar{G}$ according to the undirected degree-corrected model, and then choosing the orientation of each edge according to another matrix $\rho_{rs}$, where an edge $(u,v)$ is oriented from $u$ to $v$ with probability $\rho_{g_u,g_v}$.  Thus the total log-likelihood
is
\begin{align}
\label{eq:llh1_odc}
\like(G \mid \theta, \omega, \rho, g) = \like(\bar{G} \mid \theta, \omega, g) + \like(G \mid \bar{G}, \rho, g) \, .
\end{align}
Writing $\bar{m}_{rs} = m_{rs}+m_{sr}$ and $\kappa_r = \kappain_r+\kappaout_r$, we can set $\theta_u$ and $\omega_{rs}$ for the undirected model to their \MLEs\ as in Section~\ref{sec:dc}, giving
\begin{align}
\label{eq:llh_odc_part1}
\like(\bar{G} \mid g)=\frac{1}{2}\sum_{r,s=1}^K \bar{m}_{rs} \log \frac{\bar{m}_{rs}}{\kappa_r \kappa_s} \, .
\end{align}
The orientation term is
\begin{align}
\label{eq:llh_odc_part2}
\like(G \mid \bar{G},\rho,g)
=\sum_{rs}m_{rs} \log\rho_{rs}
=\,\frac{1}{2}\sum_{rs}(m_{rs} \log\rho_{rs}+m_{sr} \log\rho_{sr}) \, ,
\end{align}
For each $r, s$ we have $\rho_{rs}+\rho_{sr}=1$, and the \MLEs\ for $\rho$ are
\begin{align}
\hat{\rho}_{rs}=m_{rs}/ \bar{m}_{rs} \, .
\end{align}
As~\eqref{eq:llh_odc_part2} is maximized when the $\hat{\rho}_{rs}$ are near $0$ or $1$, the edge orientation term prefers highly directed inter-block connections.  Since $\hat{\rho}_{rr}=1/2$ for any $r$, it also prefers disassortative mixing, with as few connections as possible within blocks.
Substituting the \MLEs\ for $\rho$ and combining~\eqref{eq:llh_odc_part1} with~\eqref{eq:llh_odc_part2}, the total log-likelihood is
\begin{align}
\label{eq:llh_dtt}
\like(G \mid g)
=\sum_{r,s=1}^K m_{rs} \log \frac{m_{rs}}{\kappa_r\kappa_s} \, .
\end{align}

We can also view the \ODC\ as a special case of the \DDC, where we add the constraint $\thetain_u = \thetaout_u$ for all vertex $u$ (see Appendix~\ref{app:odc}). Moreover, if we set $\theta_u=1$ for all $u$, we obtain the original block model, or rather its Poisson multigraph version where each $A_{uv}$ is Poisson-distributed with mean $\omega_{g_u,g_v}$. Thus
\[
\text{SBM} \le \text{ODC} \le \text{DDC} \, ,
\]
where $A \le B$ means that model $A$ is a special case of model $B$, or that $B$ is an elaboration of $A$.  We will see below that since it is forced to explain edge orientations, the \ODC\ performs better on some networks than either the simple \SBM\ or the \DDC.

\subsection{Degree-generated block models}
\label{sec:dg-model}

Another way to utilize vertex degrees for community detection is to require the model to generate them, but according to some distribution derived from domain knowledge or an overall measurement of the network's degree distribution.  For instance, many real-world networks have a power-law degree distribution, but with parameters (such as the exponent, minimum degree, or leading constant) that vary from community to community. In that case, the degree of a vertex gives us a clue as to its block membership.  Our \emph{degree-generated} (\DG) block models allow heavy-tailed degree distributions, unlike the simple block model, while taking advantage of vertex degrees to help it classify the vertices, unlike the degree-corrected model of Karrer and Newman.


To maintain the tractability of the model, we do not generate the degrees directly.  Instead, we generate the $\theta$ parameters of one of the degree-corrected block models discussed above, and use them to generate a random multigraph.  Specifically, each $\theta_u$ is generated independently according to some distribution whose parameters $\psi$ depend on the block $g_u$ to which $u$ belongs.
Thus the \DG\ model is a \emph{hierarchical model}, which extends the previous degree-corrected block models by adding a degree generation stage on top, treating the $\theta$s as generated by the block assignment $g$ and the parameters $\psi$ rather than as parameters.

We can apply this approach to the undirected, directed, or oriented versions of the degree-corrected model; at the risk of drowning the reader in acronyms, we denote these \DGDC, \DGDDC, and \DGODC.  In each case, the total log-likelihood of a graph $G$ is
\[
\log P(G \mid \psi, \omega, g) = \log \int \dtheta \,P(G \mid \theta, \omega, g) \,P(\theta \mid \psi, g) \, ,
\]
where
\[
P(\theta \mid \psi, g) = \prod_u P(\theta_u \mid \psi_{g_u}) \, .
\]
For the directed models, we use $\theta_u$ as a shorthand here for $\thetain_u$ and $\thetaout_u$.

As in many hierarchical models, computing this integral appears to be difficult, except when $P(\theta \mid \psi)$ has the form of a conjugate prior such as the Gamma distribution (see Appendix C).  We approximate it by assuming that it is dominated by the most-likely value of $\theta$,
\[
\log P(G \mid \psi, \omega, g) \approx \log P(G \mid \thetahat, \omega, g) + \log P(\thetahat \mid \psi, g) \, .
\]
However, even determining $\thetahat$ is challenging where $P(\theta \mid \psi)$ is, say, a power law with a minimum-degree cutoff.  Thus we make a further approximation, setting $\thetahat$ just by maximizing the block model term $\log P(G \mid \thetahat, \omega, g)$ as we did before, using~\eqref{eq:mle_doi} or the analogous equations for the \DC\ or \ODC.
In essence, these approximations treat $P(\thetahat \mid \psi, g)$ as a penalty term, imposing a prior probability on the degree distribution of each community with hyperparameters $\psi$.  This leaves the door open for community structures that might not be as good a fit to the edges, but compensate with a much better fit to the degrees.

We can either treat the degree-generating parameters $\psi$ as fixed (say, if they are predicted by a theoretical model of network growth~\cite{BAmodel,bauke-etal,MGN06}) or infer them by finding the $\psihat$ that maximizes $P(\thetahat \mid \psi)$.  For instance, suppose the $\theta_u$ in block $g_u=r$ are distributed as a continuous power law with a lower cutoff $\thetaminr$.  Specifically, let the parameters in each block $r$ be $\psi_r = (\alpha_r, \beta_r, \thetaminr)$, and
\[
P(\theta_u \mid \psi_r) =
\begin{cases}
\beta_r & \theta_u=0 \\
0 & 0 < \theta_u < \thetaminr \\
\frac{(1-\beta_r)(\alpha-1)}{\thetaminr} \left( \frac{\theta_u}{\thetaminr} \right)^{-\alpha_r} & \theta_u \ge \thetaminr \, .
 \end{cases}
\]
In the directed case, we have
$\psi_r^{\textrm{in}} = (\alpha_r^{\textrm{in}}, \beta_r^{\textrm{in}}, \thetaminr^{\textrm{in}})$ and
$\psi_r^{\textrm{out}} = (\alpha_r^{\textrm{out}}, \beta_r^{\textrm{out}}, \thetaminr^{\textrm{out}})$.  Allowing $\beta_r^{\textrm{out}}$ to be nonzero, for instance, lets us directly include nodes with no outgoing neighbors; we find this useful in some networks.  Alternately, we can choose $(\thetain_u,\thetaout_u)$ from some joint distribution, allowing in- and out-degrees to be correlated in various ways.

We fix $\thetaminr = 1$.  Given the degrees and the block assignment, the \MLEs\ for $\alpha_r$ and $\beta_r$ are as follows.  Let $Y_r = \{ u: g_u=r \text{ and } \theta_u \ne 0 \}$, and let $y_r = |Y_r|$.  Then the most-likely exponent of the power law is~\cite{CLAUSET:2007}
\begin{equation}
\label{eq:clauset}
\hat{\alpha}_r = 1+ y_r \left\slash \sum_{u \in Y_r} \ln \theta_i \right. \, .
\end{equation}
The \MLE\ for $\hat{\beta_r}$ is simply the fraction of vertices in block $r$ with degree zero.

\section{Experiments on synthetic networks}

In this section, we describe experiments on various synthetic networks.  First, we generated undirected networks according to the \DGDC\ model, with two blocks or communities of equal size $N/2$.  In order to confound the block model as much as possible, we deliberately designed these networks so that the two blocks have the same average degree.  The degree distribution in block $1$ is a power law with exponent $\alpha=1.7$, with an upper bound of $1850$, so that the average degree is $20$.  The degree distribution in block $2$ it is Poisson, also with mean $20$.  As described in Appendix D, the upper bound on the power law is larger than any degree actually appearing in the network; it really just changes the normalizing constant of the power law, and
the \MLE\ for $\alpha$ can still be calculated using~\eqref{eq:clauset}.
We assume the algorithm knows that one block has a power law degree distribution and the other is Poisson, but we force it to infer the parameters of these distributions.

As in~\cite{KARRER:2011:ID106}, we use a parameter $\lambda$ to interpolate linearly between a fully random network with no community structure and a ``planted'' one where the communities are completely separated.  Thus
\[
\omega_{rs} = \lambda \omega_{rs}^{\text{planted}}+(1-\lambda)\omega_{rs}^{\text{random}}
\]
where
\[
\omega_{rs}^{\text{random}} = \frac{\kappa_r \kappa_s}{2M}
\, , \;
\omega^{\text{planted}}= \left( \begin{array}{cc}
\kappa_1 & 0 \\
0 & \kappa_2 \end{array} \right) \, .
\]
We inferred the community structure with various models. We ran the Kernighan-Lin (KL) heuristic first to find a local optimum~\cite{KARRER:2011:ID106}, and then ran the heat-bath \MCMC\ algorithm with fixed number of iterations to further refine it if ever possible.
We initialized each run with a random block assignment; to test the stability of the models, we also tried initializing them with the correct block assignment.  Since isolated vertices don't participate in the community structure, giving us little or no basis on which we can classify them, we remove them and focus on the giant component. For $\lambda=1$, where the community structure is purely the ``planted" one, we kept two giant components, one in each community.

We measured accuracy by the normalized mutual information (\NMI)~\cite{Danon2005} between the most-likely block assignment found by the model and the correct assignment.  To make this more concrete, if there are two blocks of equal size and $95\%$ of the vertices in each block are labeled correctly, the \NMI\ is $0.714$.  If $90\%$ in each group are labeled correctly, the \NMI\ is $0.531$.  For groups of unequal size, the \NMI\ is a better measure of accuracy than the fraction of vertices labeled correctly, since one can make this fraction fairly large simply by assigning every vertex to the larger group.

\begin{figure}[]
\begin{center}
\includegraphics[scale=0.24]{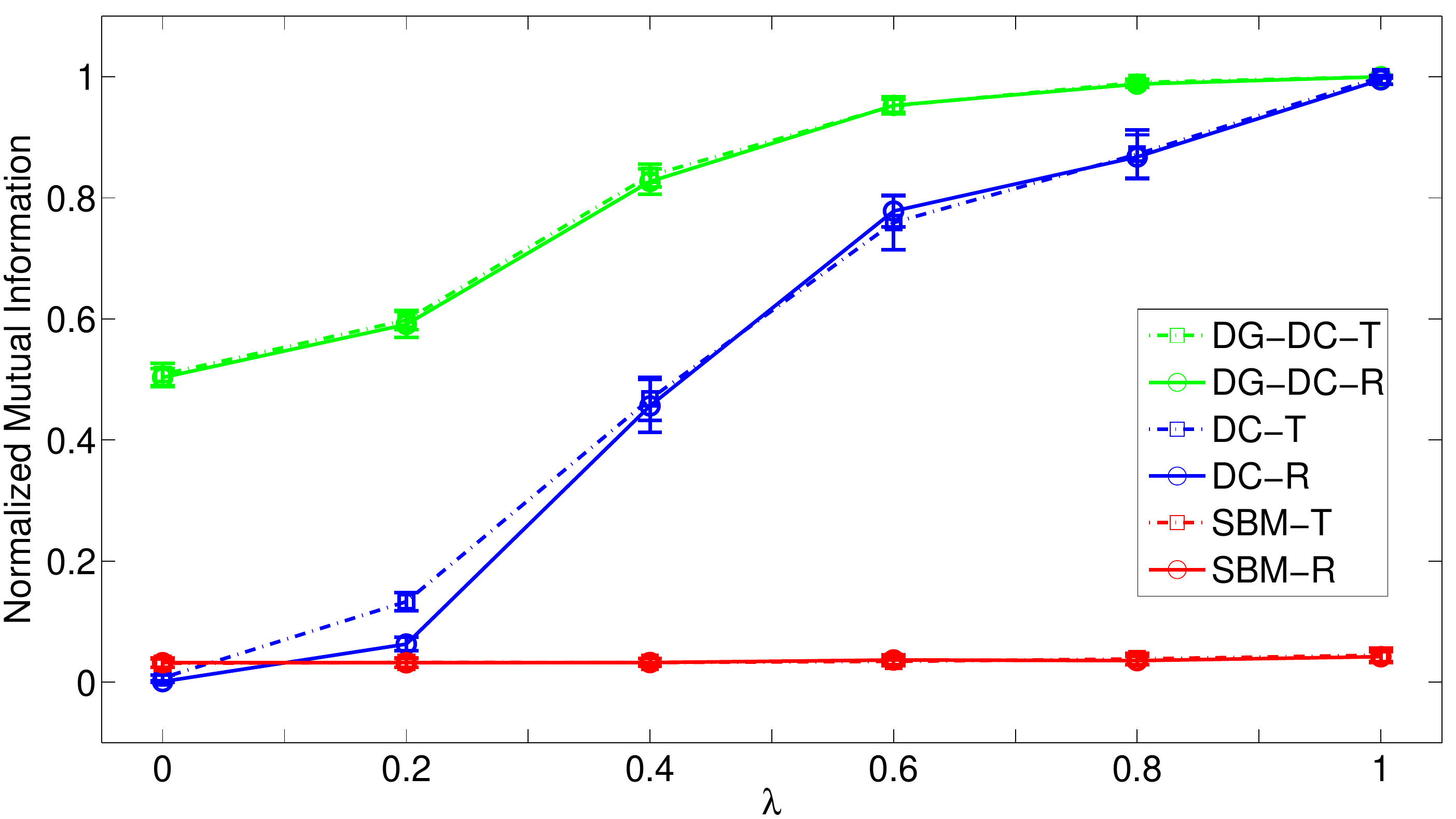}
\end{center}
\caption{Tests on synthetic networks generated by the \DGDC\ model. Each point is based on $30$ randomly generated networks with $N=2400$.  For each network and each model, we choose the best result from $10$ independent runs, initialized either with random assignments (the suffix $R$) or the true block assignment (the suffix $T$).  Each run consisted of the KL-heuristic followed by $10^6$ \MCMC\ steps. Our degree-generated (\DG) block model performs much better on these networks than the degree-corrected (\DC) model. The non-degree-corrected (\SBM) model doesn't work at all.
}
\label{fig:Syn_DC}
\end{figure}

As shown in Fig.~\ref{fig:Syn_DC}, \DGDC\ works very well even for small $\lambda$.  This is because it can classify most of the vertices simply based on their degrees; if $d_u$ is far from $20$, for instance, then $u$ is probably in block $1$.  As $\lambda$ increases, it uses the connections between communities as well, giving near-perfect accuracy for $\lambda \ge 0.6$.  It does equally well whether its initial assignment is correct or random.

The \DC\ model, in contrast, is unable to use the vertex degrees, and has accuracy near zero (i.e., not much better than a random block assignment) for $\lambda \le 0.2$.  Like the \SBM~\cite{decelle_2011_2,decelle_2011}, it may have a phase transition at a critical value of $\lambda$ below which the community structure is undetectable.  Initializing it with the correct assignment helps somewhat at these values of $\lambda$, but even then it settles on an assignment far from the correct one.

The original stochastic block model (\SBM), which doesn't correct the degrees, separates vertices with high degrees from vertices with low degrees. Thus it cannot find the correct group structure even for large $\lambda$.  Our synthetic tests are designed to have a broad degree distribution in block 1, and thus make \SBM\ fail.  Note that if the degree distribution in block 1 is a power-law with a larger exponent $\alpha$, then most of the degrees will be much lower than 20, in which case \SBM\ works reasonably well.

Next, we generated directed networks according to the \DGDDC\ model.  We again have two blocks of equal size, with degree distributions similar to the undirected networks tested above.  In block $1$, both out- and in-degrees are power-law distributed with $\alpha=1.7$, with an upper bound $1850$ so that the expected degree is $20$.  In block $2$, both out- and in-degrees are Poisson-distributed with mean $20$. To test our oriented and directed models, we interpolate between a random network $\omega_{rs}^{\text{random}}=\kappa_r\kappa_s/4M$ and a planted network with completely asymmetric connections between the blocks,
\begin{equation}
\omega^{\text{planted}}= \left( \begin{array}{cc}
(\kappa_1-\omega_{12})/2 & \omega_{12} \\
0 & (\kappa_2-\omega_{12})/2 \end{array} \right) \, ,
\end{equation}
where
$\omega_{12}\le\text{min}(\kappa_1,\kappa_2)$. We choose $\omega_{12}=\frac{1}{2}\text{min}(\kappa_1,\kappa_2)$.

\begin{figure}[]
\centering
\subfigure{
\includegraphics[scale=0.20]{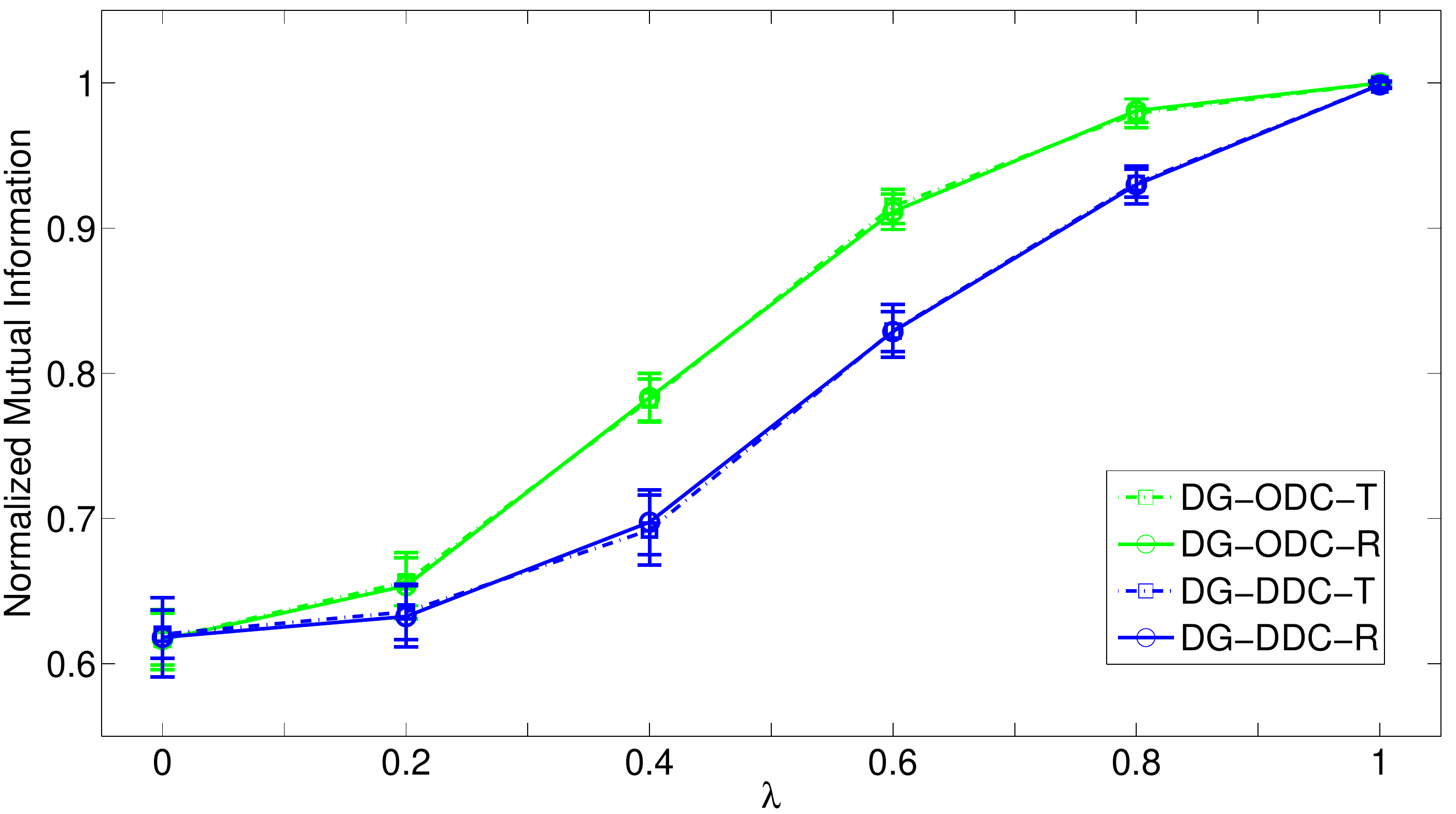}
\label{fig:Syn_ODC_DG}
}
\subfigure{
\includegraphics[scale=0.20]{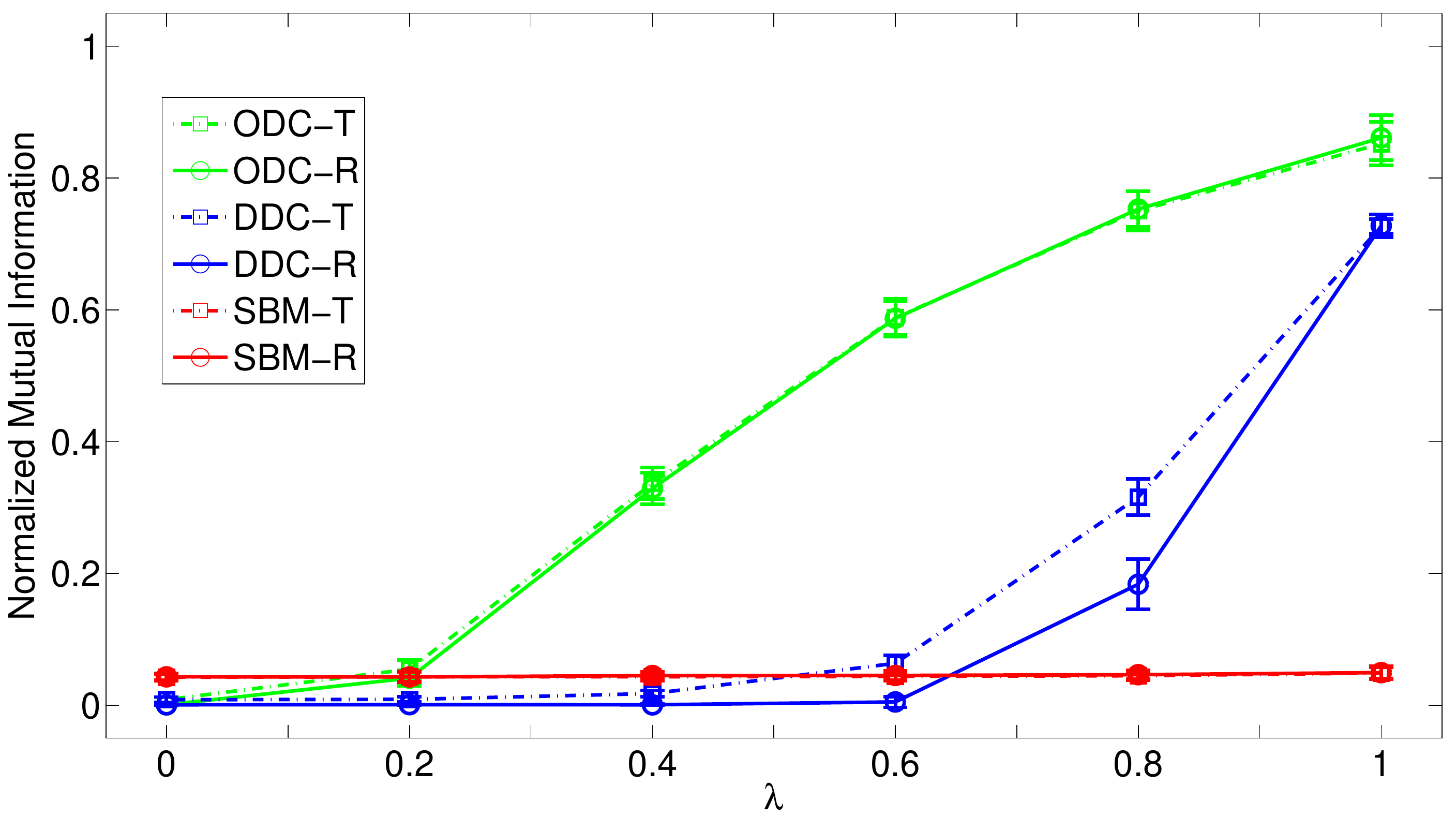}
\label{fig:Syn_ODC_noDG}
}
\caption{Tests on synthetic directed networks with $N=2400$.  Left, \DGODC\ and \DGDDC; right, \ODC\ and \DDC.  The degree-generated models again perform very well even for small $\lambda$, since they can use in- and out-degrees to classify the vertices.  \ODC\ performs significantly better than \DDC\ for $\lambda \ge 0.4$, since it can use the edge orientations to distinguish the two blocks.  The number of networks, runs, and \MCMC\ steps per run are as in Fig.~\ref{fig:Syn_DC}.}
\label{fig:Syn_ODC}
\end{figure}

As Fig.~\ref{fig:Syn_ODC} shows, \DGODC\ and \DGDDC\ have very similar performance at the extremes where $\lambda=0$ and $1$.  However, \DGODC\ works better than \DGDDC\ for other $\lambda$, and both of them achieve much better accuracy than the \ODC\ or \DDC\ models.  As in Fig.~\ref{fig:Syn_DC}, the degree-generated models can achieve a high accuracy based simply on the vertex degrees, and as $\lambda$ grows they leverage this information further to achieve near-perfect accuracy for $\lambda \ge 0.8$.

Among the non-degree-corrected models, \ODC\ performs significantly better than \DDC\ for $\lambda \ge 0.4$.  Edges are more likely to point from block $1$ to block $2$ than vice versa, and \ODC\ can take advantage of this information while \DDC\ cannot.  As we will see in the next section, \ODC\ performs well on some real-world networks for precisely this reason.



\section{Experiments on real networks}

In this section, we describe experiments on three word adjacency networks in which vertices are separated into two blocks: adjectives and nouns.  The first network consists of common words in Dickens' novel \emph{David Copperfield}~\cite{NEWMAN:2006:ID174}. The other two are formed by adjectives and nouns in the Brown corpus, which is a tagged corpus of present-day edited American English across various categories, including news, novels, documents, and many others~\cite{brownCorpus}.  We build two different networks from the Brown corpus.  The smaller one contains words in the News category (45 archives) that appeared at least 10 times; the larger one contains all the adjectives and nouns in the giant component of the entire corpus.

We considered both the simple version of these networks where $A_{uv}=1$ if $u$ and $v$ ever occur together in that order, and the multigraph version where $A_{uv} \ge 0$ is the number of times they occur together.  The sizes, block sizes, and number of edges of these networks are shown in Table~\ref{tab:word_net_stats}.  In ``News" and ``Brown", the block sizes are quite different, with more nouns than adjectives.  As discussed above, the \NMI\ is a better measure of accuracy than the fraction of vertices labeled correctly, since we could make the latter fairly large by labeling everything a noun.

In each network, both blocks have heavy-tailed in- and out-degree distributions.  The connections between them are disassortative and highly asymmetric: since in English adjectives precede nouns more often than they follow them, and more often than adjectives precede adjectives or nouns precede nouns, $\omega_{12}$ is roughly $10$ times larger than $\omega_{21}$, and $\omega_{12}$ is larger than either $\omega_{11}$ or $\omega_{22}$.  The $\omega$ for each network corresponding to the correct block assignment (according to the stochastic block model) is shown in Table~\ref{tab:words_connection_matrix}.



\begin{table}[]
  \centering
    \caption{Basic statistics of the three word adjacency networks. $S$ and $M$ denote the simple and multigraph versions respectively.}
    \begin{tabular}{l|l|l|l|l|l}
    \addlinespace
    \toprule
    Network & \#words & \#adjective & \#noun & \#edges (S) & \#edges (M)\\
    \midrule
    David & 112  & 57  & 55 & 569 & 1494\\
    News & 376  & 91  & 285  & 1389 & 2411\\
    Brown & 23258  & 6235  & 17023 & 66734 & 88930\\
    \bottomrule
    \end{tabular}%
  \label{tab:word_net_stats}%
\end{table}%

\vspace{-0.2in}

\begin{table}[]
  \centering
    \caption{The matrices $\omega_{rs}=m_{rs}/(n_r n_s)$ for the most-likely block assignment according to the stochastic block model.}
    \begin{tabular}{cc|cc|cc|cc|cc|cc}
    \addlinespace
    \toprule
          \multicolumn{2}{c}{David(S)} & \multicolumn{2}{c}{David(M)} & \multicolumn{2}{c}{News(S)} & \multicolumn{2}{c}{News(M)} & \multicolumn{2}{c}{Brown(S)} & \multicolumn{2}{c}{Brown(M)}\\
    \midrule
    0.039\;   & 0.118\;   & 0.080\;  & 0.358\; & 0.010\;   & 0.015\;   & 0.012\;   & 0.028\; & 9.1e-05\;   & 3.4e-04\; & 1.1e-04\; & 4.4e-04\\
    0.018\;   & 0.006\;  & 0.025\;  & 0.011\; & 0.002\;   & 0.010\;  & 0.003\;   & 0.019\;  & 2.0e-05\;   & 8.8e-05\; & 2.4e-05\; & 1.2e-04\\
    \bottomrule
    \end{tabular}%
  \label{tab:words_connection_matrix}%
\end{table}%

\subsection{Performance of oriented and degree-corrected models}

Table~\ref{tab:words_clustering_results_randInit} compares the performance of non-degree-generated block models, including \SBM, \DC, \ODC, and \DDC.  When applying \DC, we ignore the edge orientations, and treat the graph or multigraph as undirected
(note that the resulting network may contain multi-edges even though the directed one doesn't).

In our experiments, we started with a random initial block assignment, ran the Kernighan-Lin (KL) heuristic to find a local optimum~\cite{KARRER:2011:ID106}, and then ran the heat-bath \MCMC\ algorithm.
We also tested a naive heuristic \mbox{(NH)} which simply labels a vertex $v$ as an adjective if $\dout_v > \din_v$, and a noun if $\din_v > \dout_v$.  If $\dout_v=\din_v$, \NH\ labels $v$ randomly with equal probabilities.

\begin{table}[]
  \centering
  \caption{
For each model and each network, we pick the block assignment with highest likelihood and compute its \NMI\ with the correct block assignment.  Each run consisted of the KL-heuristic, starting with a random block assignment, followed by $10^6$ \MCMC\ steps.  The results for ``David'' and ``News'' are based on 100 independent runs; for ``Brown'', 50 runs are executed.  The best \NMI\ for each network is shown in bold.}
    \begin{tabular}{ccccccc}
    \addlinespace
    \toprule
          & \;\;\;\mbox{David(S)}\;\; & \;\;\;\mbox{David(M)}\;\; & \;\;\;\mbox{News(S)}\;\;\; & \;\;\;\mbox{News(M)}\;\; & \;\;\mbox{Brown(S)}\;\; & \;\mbox{Brown(M)}     \\
    \midrule
    SBM\;\;	& .423 & .051 & .006 & .018 & .001 & 7e-04   \\
    DC\;\;	& \textbf{.566} & \textbf{.568} & .084 & .083 & .020 & .015   \\
    ODC\;\;	& .462 & .470 & .084 & .029 & \textbf{.311} & \textbf{.318}    \\
    DDC\;\;	& .128 & 8e-04 & .084 & .091 & .016 & .012 \\
    NH\;\;	& .395 & .449 & \textbf{.215} & \textbf{.233} & .309 & .314 \\
    \bottomrule
    \end{tabular}%
  \label{tab:words_clustering_results_randInit}%
\end{table}%

For ``David'', \DC\ and \ODC\ work fairly well, and both are better than the naive \NH.  Moreover, the mistakes they make are instructive.  There are three adjectives with out-degree zero: ``full'', ``glad'', and ``alone''.  \ODC\ mislabels these since it expects edges to point away from adjectives, while \DC\ labels them correctly by using the fact that (undirected) edges are disassortative, crossing from one block to the other.

The standard \SBM\ works well on ``David(S)'' but fails on ``David(M)'' because the degrees in the multigraph are more skewed than those in the simple one.  Finally, \DDC\ performs the worst; by correcting for in- and out-degrees separately, it loses any information that the edge orientations could provide, and even fails to notice the disassortative structure that \DC\ uses.  Thus full degree-correction in the directed case can make things worse, even when the degrees in each community are broadly distributed.

For ``Brown'', all these models fail except \ODC, although it does only slightly better than the naive \NH.  For ``News'', all these models fail, even \ODC.  Despite the degree correction, the most-likely block assignment is highly assortative, with high-degree vertices connecting to each other.
However, we found that in most runs on ``News'', \ODC\ used the edge orientations successfully to find the a block assignment close to the correct one; it found the assortative structure only occasionally.  This suggests that, even though the ``wrong'' structure has a higher likelihood, we can do much better if we know what kind of community structure to look for; in this case, disassortative and directed.


To test this hypothesis, we tried giving the models a hint about the community structure by using \NH\ to determine the initial block assignment.  We then performed the KL heuristic and the \MCMC\ algorithm as before.   As Table~\ref{tab:words_clustering_results_NHInit} shows, this hint improves \ODC's performance on ``News'' significantly; it is able to take the initial naive classification, based solely on degrees, and refine it using the network's structure.  Note that this more accurate assignment actually has lower likelihood than the one found in Table~\ref{tab:words_clustering_results_randInit} using a random initial condition---so \NH\ helps the model stay in a more accurate, but less likely, local optimum.  Starting with \NH\ improves \DC's performance on ``Brown'' somewhat, but \DC\ still ends up with an assignment less accurate than the naive one.

\begin{table}[]
  \centering
  \caption{Results using the naive \NH\ assignment as the initial condition, again followed by $10^6$ \MCMC\ steps.  This hint now lets \ODC\ outperform the other models on ``News''.}
    \begin{tabular}{ccccccc}
    \addlinespace
    \toprule
          & \;\;\;\mbox{David(S)}\;\; & \;\;\;\mbox{David(M)}\;\; & \;\;\;\mbox{News(S)}\;\;\; & \;\;\;\mbox{News(M)}\;\; & \;\;\mbox{Brown(S)}\;\; & \;\mbox{Brown(M)}     \\
    \midrule
    SBM\;\;	& .423 & .051 & .006 & .021 & .001 & 7e-04   \\
    DC\;\;	& \textbf{.566} & \textbf{.568} & .084 & .015 & .160 & .155   \\
    ODC\;\;	& .462 & .470 & \textbf{.247} & \textbf{.270} & \textbf{.311} & \textbf{.318}   \\
    DDC\;\;	& .015 & .060 & .084 & .005 & .005 & .070 \\
    NH\;\;	& .395 & .449 & .215 & .233 & .309 & .314 \\
    \bottomrule
    \end{tabular}%
  \label{tab:words_clustering_results_NHInit}%
\end{table}%

\subsection{Performance of degree-generated models}

\begin{figure}[]
\centering
\subfigure{
\includegraphics[scale=0.19]{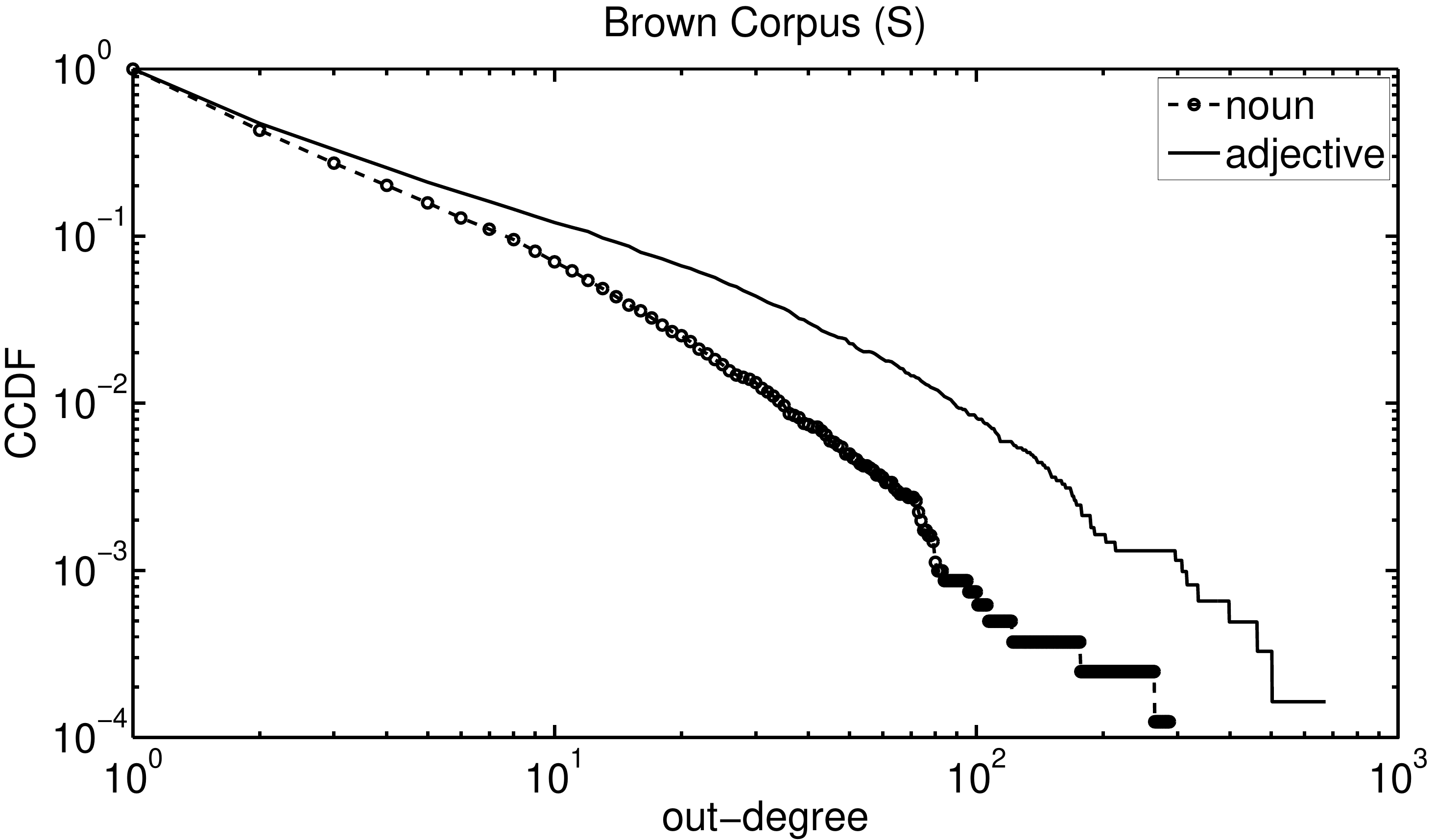}
\label{fig:Brown_S_OutDegDist}
}
\subfigure{
\includegraphics[scale=0.19]{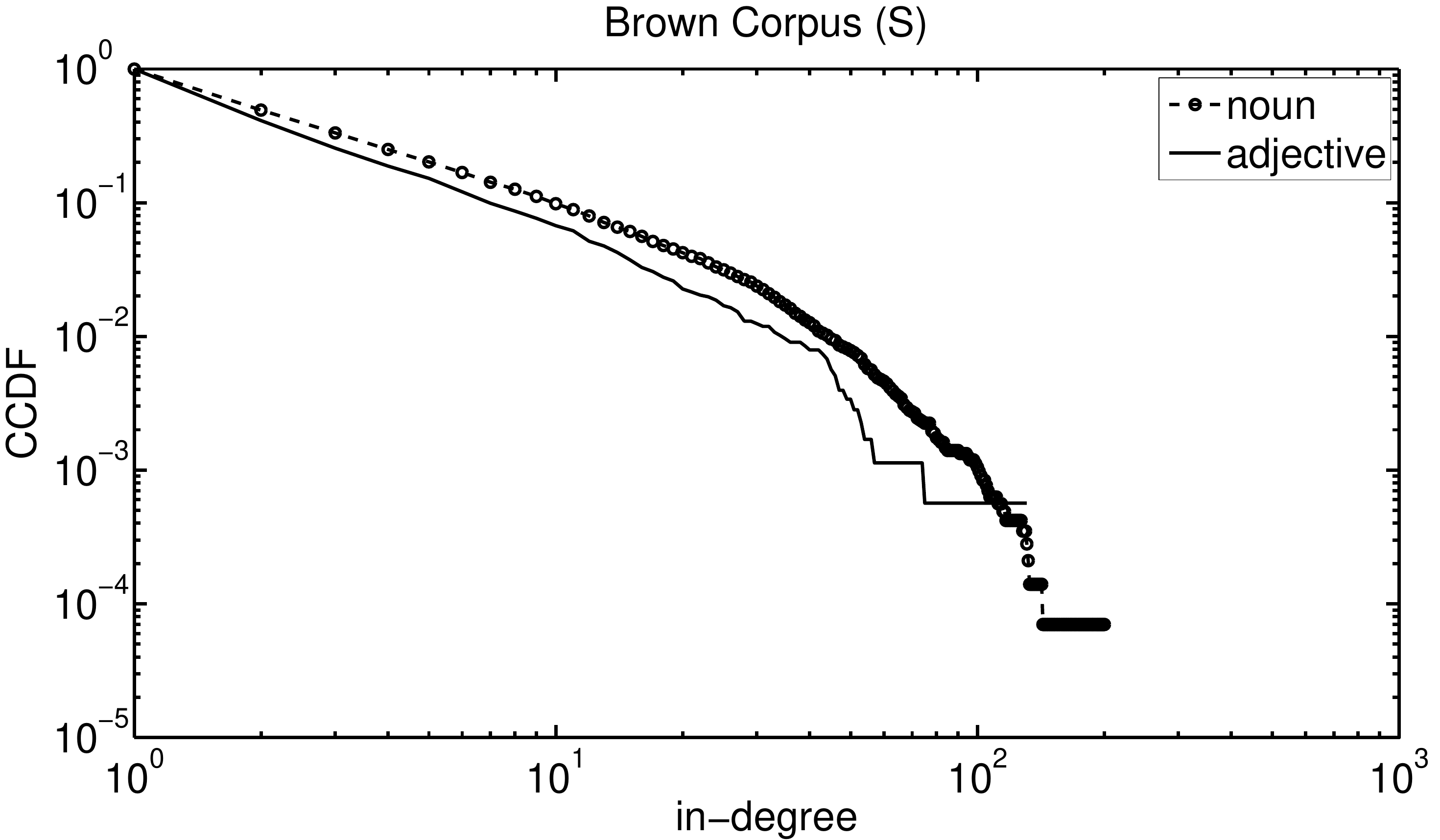}
\label{fig:Brown_S_InDegDist}
}
\subfigure{
\includegraphics[scale=0.19]{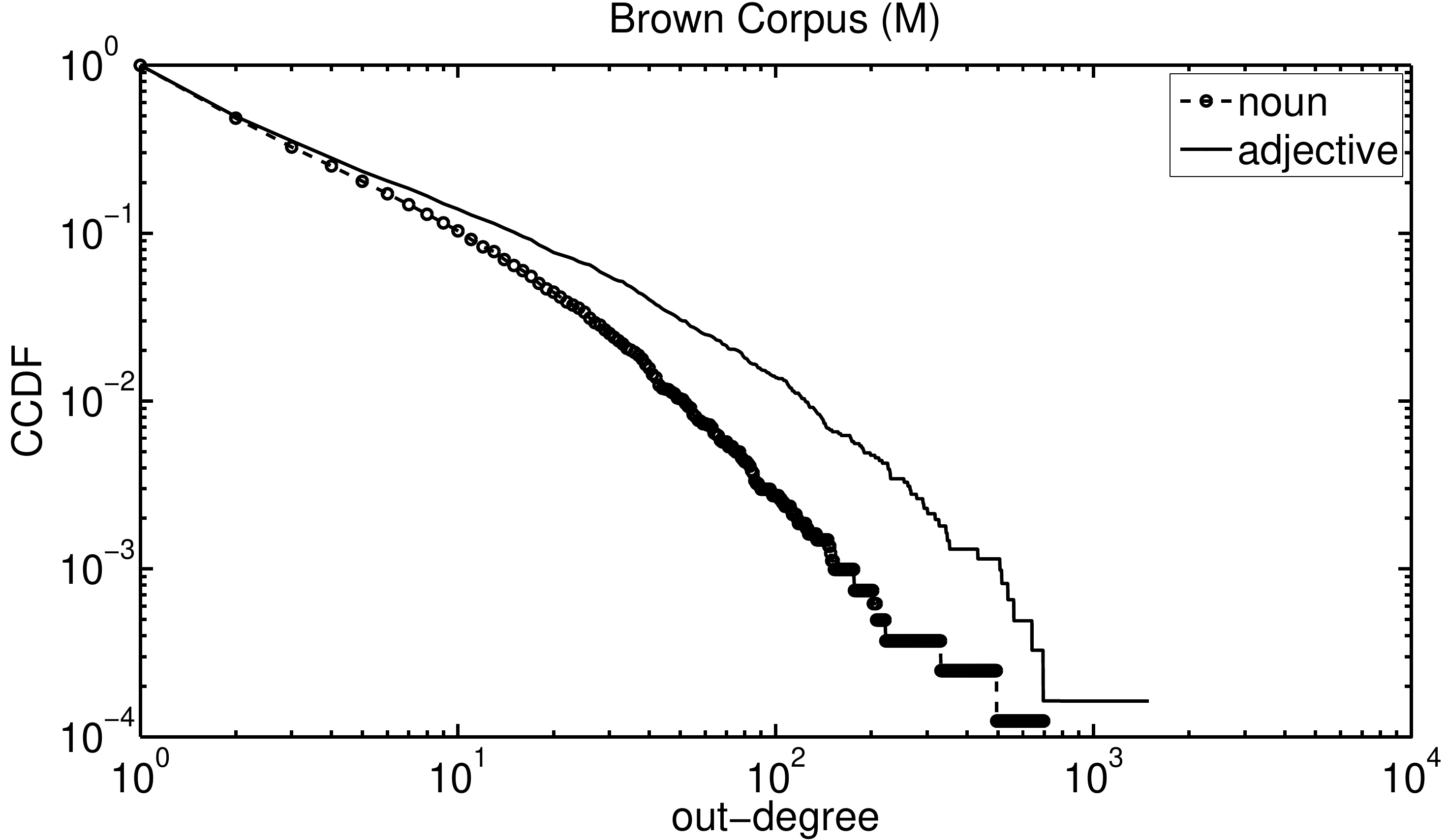}
\label{fig:Brown_M_OutDegDist}
}
\subfigure{
\includegraphics[scale=0.19]{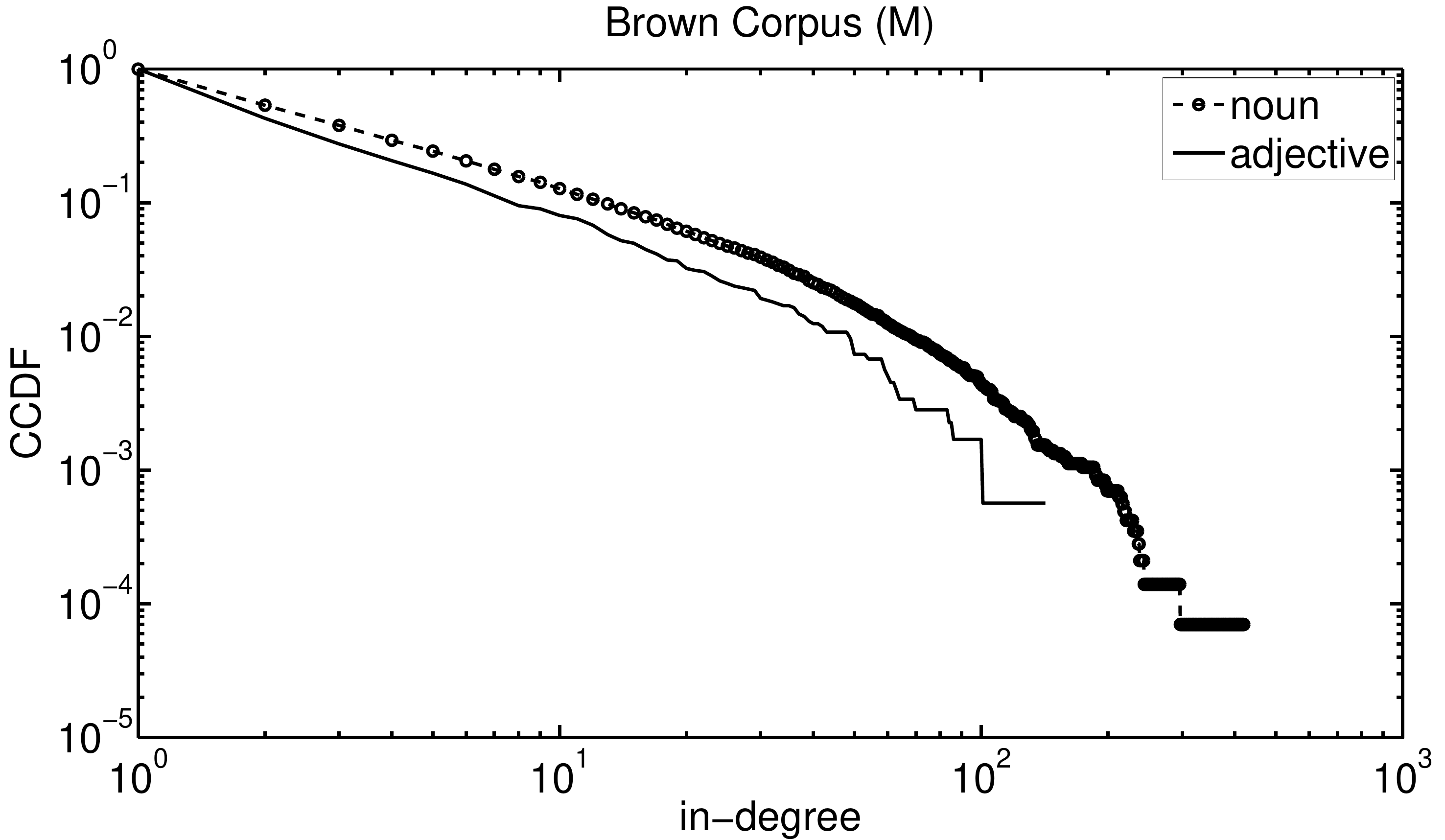}
\label{fig:Brown_M_InDegDist}
}
\caption{Degree distributions in the Brown network.}
\label{fig:brown_deg_distri}
\end{figure}

In this section, we measure the performance of degree-generated models on the Brown network, and compare them to their non-degree-generated counterparts.  As Fig.~\ref{fig:brown_deg_distri} shows, the in- and out-degree distributions in each block have heavy tails close to a power-law.  Moreover, the out-degrees of the adjectives have a heavier tail than those of the nouns, and vice versa for the in-degrees.  This is exactly the kind of difference in the degree distributions between communities that our \DG\ block models are designed to take advantage of.

Setting $\thetamin=1$, we can estimate the parameters $\alpha$ and $\beta$ for these distributions as discussed in Section~\ref{sec:dg-model}.   We show the most likely values of these parameters, given the correct assignment, in Table~\ref{tab:true-alpha-beta-brown}.
\begin{table}[]
  \centering
  \caption{MLEs for the degree generation parameters in the Brown network, given the correct assignment.}
    \begin{tabular}{lcccccccc}
    \addlinespace
    \toprule
    & \multicolumn{4}{c}{\;\;Brown(S)\;\;} & \multicolumn{4}{c}{\;\;Brown(M)\;\;}\\
    \cmidrule(r){2-5} \cmidrule(r){6-9}
        block\;     & \;$\hat{\alpha}_{\text{in}}$\; & \;$\hat{\alpha}_{\text{out}}$\; & \;$\hat{\beta}_{\text{in}}$\; & \;$ \hat{\beta}_{\text{out}}$ \; & \;$\hat{\alpha}_{\text{in}}$\; &\; $\hat{\alpha}_{\text{out}}$\; &\; $\hat{\beta}_{\text{in}}$\; &\;$\hat{\beta}_{\text{out}}$\\
    \midrule
     adjective\;\;   &2.329\; &2.629\; &0.161\; &0.527\; &2.136\; &2.326\; &0.161\; &0.527\\
     noun  &2.721\; &2.248\; &0.716\; &0.021\; &2.576\; &2.134\; &0.716\; &0.021\\
    \bottomrule
    \end{tabular}%
  \label{tab:true-alpha-beta-brown}%
\end{table}%

As Table~\ref{tab:perf-compare-brown-kl} shows, degree generation improves \DC\ and \DDC\ significantly, letting them find a good assignment as opposed to one with \NMI\ near zero.  For \ODC, the performance improvement is slight, making \DGODC\ the best model overall, but there is another effect.  We compare performance starting with the KL heuristic to performance using \MCMC\ alone.  We see that degree generation gives \ODC\ almost as much benefit as the KL heuristic does.  In other words, it speeds up the \MCMC\ optimization process, letting \ODC\ find a good assignment without the initial help of the (computationally expensive) KL heuristic.


\begin{table}[]
  \centering
  \caption{Performance of degree-generated vs.\ non-degree generated models.  KL indicates that we applied the KL heuristic and then $10^6$ \MCMC\ steps, as opposed to \MCMC\ alone.  \DG\ indicates degree generation.  Each number gives the \NMI\ for the most-likely assignment found in $50$ independent runs.  The best model is \DGODC.  Moreover, degree generation helps \ODC\ converge, providing much of the benefit of the KL heuristic while avoiding its long running time (see bold numbers).}
    \begin{tabular}{cccccccc}
    \toprule
       &   & \multicolumn{3}{c}{\;\;Brown(S)\;\;} & \multicolumn{3}{c}{\;\;Brown(M)\;\;}\\
    \cmidrule(r){3-5} \cmidrule(r){6-8}
       &   & \;\DC\;\; &\; \ODC\; &\; \DDC\;\;\; & \;\DC\;\; &\; \ODC\; &\; \DDC\;\;\;\\
    \midrule
    --\; & -- \; & .010\; & .188\; & .008\; & .007\; & .203\; & .011\;\\
    KL\; & -- \; & .020\; & .311\; & .016\; & .015\; & .318\; & .012\;\\
    --\; &DG \; & .267\; & .302\; & .213\; & .278\; & .310\; & .149\;\\
    KL\; &DG \; & .271\; & \textbf{.312}\; & .225\; & .284\; & \textbf{.320}\; & .195\;\\
    \bottomrule
    \end{tabular}%
  \label{tab:perf-compare-brown-kl}%
\end{table}%

\section{Conclusions}

Degree correction in stochastic block models provides a powerful approach to dealing with networks with inhomogeneous degree distributions.  However, in a sense it denies information to the inference process, since a generative model can only help us learn from the data that it has to generate.

We have introduced two new kinds of block models that allow for broad or heavy-tailed degree distributions, while using the degrees to help us detect communities.  The oriented degree-corrected model (\ODC) performs partial degree correction, taking the total degrees as parameters but generating edge orientations.  The degree-generated (\DG) models don't take the degrees as parameters, but assumes that they are generated according to some prior in each community.


Unlike the directed degree-corrected (\DDC) block model, which takes both in- and out-degrees as parameters, \ODC\ is able to capture and account for certain correlations between the in- and out-degrees. Simply put, for \ODC, two vertices are unlikely to be in the same community if one has high in-degree and low out-degree while another has high out-degree and low in-degree.  If the network is highly directed or asymmetric, the edge orientations can help \ODC\ find community structures that \DDC\ fails to perceive.


Our \DG\ models use degree-corrected block models as a subroutine, but impose a penalty term based on the prior likelihood of the degree distribution in each community.  They can take the (hyper)parameters of these priors as given, or infer them ``on the fly.''  \DG\ models achieve high accuracy even when the density of connections between communities is close to uniform, as we illustrated in synthetic networks for small $\lambda$.  Augmenting block models, such as the \ODC, with degree generation also appears to speed up their convergence in some cases, helping simple algorithms like \MCMC\ handle large networks without the benefit of expensive preprocessing steps like the KL heuristic.

On the other hand, the effectiveness of \DG\ depends heavily on knowing the correct form of the degree distribution in each community.  Without some prior ground truth about the block assignment, or domain-specific knowledge, finding an appropriate family of degree distributions may be difficult for some networks.



With all these variants of the block model, ranging from the ``classic'' version to degree-corrected and degree-generated variants, we now have a wide variety of tools for inferring structure in network data.  Each model will perform better on some networks and worse on others.  A better understanding of the strengths and weaknesses of each one---which kinds of structure they can see, and what kinds of structure they are blind to---will help us select the right algorithm each time we meet a new network.


\vspace{20pt}
\textbf{Acknowledgments.} We are grateful to Terran Lane, Ben Edwards, Aaron Clauset, and Mark Newman for helpful conversations, and to the McDonnell Foundation for their support.
\bibliography{dg-paper}
\bibliographystyle{plain}

\appendix

\section{Maximum Likelihood Estimators for
the directed degree-corrected (\DDC) block model}
\label{app:ddc}

We maximize the log-likelihood function~\eqref{eq:llh1_doi},
{\begin{align}
\log
P(G\mid \theta,\omega,g)
&= \sum_u(\dout_u\log\thetaout_u +\din_u \log \thetain_u) \nonumber \\
&+ \sum_{rs}(m_{rs} \log \omega_{rs} - \kappaout_r \kappain_s \omega_{rs}) \, ,
\end{align}
where we have imposed the constraints on the $\theta$ parameters
\begin{equation}
\label{eq:ddc-constraints}
\sum_{u:g_u=r} \thetaout_u=\kappaout_r
\quad \text{and} \quad
\sum_{u:g_u=r} \thetain_u=\kappain_r  \, .
\end{equation}
For each block $r$, we associate Lagrange multipliers $\lambdaout_r, \lambdain_r$ with these constraints.  For each vertex $u$, taking the partial derivative of the log-likelihood with respect to $\thetaout_u$ and $\thetain_u$ gives
%
\begin{equation}
\frac{\dout_u}{\thetaout_u} = \lambdaout_{g_u}
\quad \text{and} \quad
\frac{\din_u}{\thetain_u} = \lambdain_{g_u} \, .
\end{equation}
To satisfy the constraints~\eqref{eq:ddc-constraints}, we take $\lambdaout_r = \lambdain_r = 1$ for all $r$, so that
\[
\thetahatout_u = \dout_u
\quad \text{and} \quad
\thetahatin_u = \din_u \, .
\]
Setting the partial derivative of the log-likelihood function with respect to $\omega_{rs}$ to zero then gives
\[
\hat{\omega}_{rs} = \frac{m_{rs}}{\kappaout_r\kappain_s} \, .
\]

\section{Another view of the \ODC\ model}
\label{app:odc}

Here we show that the oriented degree-corrected (\ODC) model is a special case of the directed degree-corrected (\DDC) model.  Recall that the ODC model first generates an undirected graph according to the DC model with parameters $\theta_u$ and $\omega_{rs}$, and then orients each edge $(u,v)$ from $u$ to $v$ with probability $\rho_{g_u,g_v}$.  The number of directed edges from $u$ to $v$ is then Poisson-distributed as
\[
A_{uv} \sim \Poi(\theta_u \theta_v \omega_{g_u,g_v} \rho_{g_u,g_v} ) \, .
\]
But if we write
\[
\omega'_{rs} = \omega_{rs} \rho_{rs} \, ,
\]
then
\[
A_{uv} \sim \Poi(\theta_u \theta_v \omega'_{g_u,g_v} ) \, .
\]
Thus ODC is the special case of DDC where $\thetain_u = \thetaout_u = \theta_u$ for all vertices $u$.

For completeness, we check that the two models correspond when we set these parameters equal to their MLEs.  We impose the constraint $\sum_{u:g_u=r}\theta_u=\kappa_r = \kappaout_r + \kappain_r$ for all blocks $r$.
Ignoring constants, the log-likelihood is then
\begin{equation}
\label{eq:llh_odc_withparams}
\like(G\mid \theta,\omega',g)
= \sum_u d_u \log \theta_u + \sum_{rs} (m_{rs} \log \omega'_{rs} - \kappa_r \kappa_s \omega'_{rs}) \, ,
\end{equation}
where $d_u = \dout_u + \din_u$.  The MLEs for $\theta_u$  and $\omega'_{rs}$ are then
\begin{align}
\label{eq:mle_odc}
\hat{\theta}_u=d_u \, , \quad \hat{\omega}'_{rs} = \frac{m_{rs}}{\kappa_r \kappa_s} \, .
\end{align}
Thus $\hat{\omega}'_{rs} = \hat{\omega}_{rs} \hat{\rho}_{rs}$ where
\[
\hat{\omega}_{rs} = \frac{\bar{m}_{rs}}{\kappa_r \kappa_s}
\quad \text{and} \quad
\hat{\rho}_{rs} = \frac{m_{rs}}{\bar{m}_{rs}} \, ,
\]
recovering~\eqref{eq:llh_dtt}.

\section{Bayesian estimation for \DG\ models}

Bayesian inference focuses on posterior distributions of parameters rather than on point estimates.  In hierarchical models like \DGDDC, the full Bayesian posterior of the $\theta$ parameters (omitting the other parameters $g$ and $\omega$) is
\begin{align*}
P(\theta\mid G) &= \int P(\theta\mid G,\psi) \,P(\psi\mid G) \,\dpsi \, .
\end{align*}
Here we employ the Empirical Bayesian method, and use point estimates for the hyperparameters $\psi$, namely their MLEs $\hat{\psi}$,
\begin{align}
\label{eq:bayes_psi}
\hat{\psi} &= \argmax_{\psi} P(G \mid \psi) \nonumber\\
	   &= \argmax_{\psi} \int P(G\mid \theta, \psi) \,P(\theta \mid \psi) \,\dtheta \, .
\end{align}
With this approximation we have
\begin{align}
\label{eq:bayes_theta}
P(\theta\mid G) &\approx P(\theta\mid G,\hat{\psi}) \nonumber\\
		&= \dfrac{P(G\mid \theta)\,P(\theta\mid \hat{\psi})}{P(G\mid \hat{\psi})} \nonumber\\
		&= \dfrac{P(G\mid \theta)\,P(\theta\mid \hat{\psi})}{\int P(G\mid \theta, \hat{\psi})\,P(\theta\mid \hat{\psi})\,\dtheta} \, ,
\end{align}
where we used Bayes' rule in the second line.

Computing the posterior $P(\theta\mid G)$ is usually difficult, as the integral in the denominator of \eqref{eq:bayes_theta} is often intractable. However, with a clever choice of the prior distribution $P(\theta\mid \psi)$, we can work out an analytic solution. It is called the \emph{conjugate prior} of the likelihood term.  We focus here on \DGDDC; the calculations for other degree-generated models are similar.

Say that a random variable $X$ is Gamma-distributed with parameters $\alpha, \beta$, and write $X \sim \Gamma(\alpha,\beta)$, if its probability distribution is
\[
f(x;\alpha,\beta) = \frac{\beta^\alpha}{\Gamma(\alpha)} \,x^{\alpha-1} \,\e^{-\beta x} \, .
\]
In \DGDDC, the likelihood~\eqref{eq:lh_doi} can be written (where we have plugged in the MLEs for $\omega$, and substituted $\kappaout_r = \sum_{u:g_u=r}\dout_u$)
\begin{equation}
\label{eq:dg-gamma}
P(G\mid \thetaout)
= \dfrac{\prod_u(\thetain)^{\din_u}\prod_{rs}\omega_{rs}^{m_{rs}}}{\prod_{uv}A_{uv}!}
\prod_u (\thetaout_u)^{\dout_u} \exp\!\left( -\thetaout_u
\right) \, .
\end{equation}

If we assume that the $\thetain$ and $\thetaout$ for each $u$ are independent, this is proportional to a product of Gamma distributions with parameters $\alpha=\dout_u+1$ and $\beta = 1$ for each $\thetaout_u$.

A natural conjugate prior for Gamma distributions is the Gamma distribution itself.  Let the hyperparameters $\psiout_r$ for each block $r$ consist of a pair $(\alphaout_r, \betaout_r)$, and consider the prior
\[
\thetaout_u \sim \Gamma(\alphaout_{g_u}, \betaout_{g_u}) \, .
\]
That is,
\[
P(\thetaout_u \mid \psiout_{g_u})
= \dfrac{(\betaout_{g_u})^{\alphaout_{g_u}}}{\Gamma(\alphaout_{g_u})}(\thetaout_u)^{\alphaout_{g_u}-1}\exp(-\betaout_{g_u}\thetaout_u) \, ,
\]
Multiplying this prior by the likelihood~\eqref{eq:dg-gamma} stays within the family of Gamma distributions, and simply updates the parameters:
\begin{align*}
P(\thetaout_u \mid G)
& \propto P(\thetaout_u \mid \psi^{\mathrm{out}}_{g_u}) \,P(G\mid \thetaout) \\
& \propto (\thetaout_u)^{\alphaout_{g_u}+\dout_u-1} \exp\!\left( -\thetaout_u \left( \betaout_{g_u} + 1
\right) \right) \, .
\end{align*}
Thus the posterior distribution is
\[
\thetaout_u \sim \Gamma\!\left(\alphaout_{g_u}+\dout_u, \betaout_{g_u}+
1
\right) \, .
\]
Note that if we use a uninformative prior, i.e., in the limit $\alphaout_{g_u}=1$ and $\betaout_{g_u}=0$, the Gamma prior reduces to a uniform prior.
The maximum a posteriori (MAP) estimate of $\thetaout_u$ is
\begin{align}
\label{eq:bayes_thetau}
 \thetahatout_u  = \dout_u \, ,
\end{align}
and similarly for $\thetain_u$, just as we obtained for the MLEs in~\eqref{eq:mle_doi}.

However, our goal is to integrate over $\theta$, not focus on its MAP estimate.  So let us continue the Bayesian analysis. Assuming the $\theta$ parameters are independent, then their joint posterior is simply a product of their individual posteriors
\begin{align}
 &P(\theta|G)  = \prod_uP(\thetaout_u|G)P(\thetain_u|G)\nonumber\\
 =&\prod_u f\!\left(\thetaout_u;\alphaout_{g_u}+\dout_u, \betaout_{g_u} +
 1
 \right)
 f\!\left(\thetain_u;\alphain_{g_u}+\din_u, \betain_{g_u} + 1
 \right)\,.
\end{align}
Then we can calculate the integral in \eqref{eq:bayes_psi} and \eqref{eq:bayes_theta} by the simple algebra:
\begin{align}
\label{eq:bayes_integral}
&\int P(G\mid \theta, \psi)P(\theta\mid \psi) \,\dtheta = \dfrac{P(G\mid \theta)P(\theta\mid \psi)}{P(\theta\mid G)}\\
	=& \dfrac{\prod_u f(\thetaout_u;\dout_u+1, 1
	)\,
    f(\thetain_u;\din_u+1, 1
	)\,
	f(\thetaout_u;\alphaout_{g_u}, \betaout_{g_u})\,
    f(\thetain_u;\alphain_{g_u}, \betain_{g_u})
    }
	  {\prod_u f\!\left(\thetaout_u;\alphaout_{g_u}+\dout_u, \betaout_{g_u} + 1
	  \right)
 f\!\left(\thetain_u;\alphain_{g_u}+ \din_u, \betain_{g_u} +
 1
 \right)}\nonumber\\
  =& \frac{\prod_u
 {\betaout_{g_u}}^{\alphaout_{g_u}}
 {\betain_{g_u}}^{\alphain_{g_u}}
 \Gamma\!\left(\alphaout_{g_u}+\dout_u\right)\Gamma\!\left(\alphain_{g_u}+\din_u\right)
 }
 {
 \prod_u \left(\betaout_{g_u}+
 1
 \right)^{\alphaout_{g_u}+\dout_u}\left(\betain_{g_u} +
 1
 \right)^{\alphain_{g_u}+\din_u}
 \Gamma\!\left(\dout_u+1\right)\Gamma\!\left(\din_u+1\right)\Gamma\!\left(\alphaout_{g_u}\right)\Gamma\!\left(\alphain_{g_u}\right)
 }\nonumber\, .
\end{align}
Now that the dependence of the numerator and denominator on $\theta$ has cancelled out, the integral is a function only of the hyperparameters $\psi$, making it possible to do the point estimate of $\psi$ in \eqref{eq:bayes_psi}. In our case, optimizing for $\hat{\psi}$ requires some numeric techniques, but it is nonetheless doable.

Empirical Bayesian solution not only gives better approximation to the original problem, it also make it possible to integrate prior knowledge if available. On top of that, because the posterior is now a direct function of the hyperparameters $\psi$, we no longer have to worry about the Poisson noise when estimating $\psi$ indirectly from degrees.

On a final note, the above result only holds for Gamma priors.  With any other prior, the integral may not be this simple.

\section{Power-law distribution with upper bound}

In this section, we show that imposing an upper bound on our power-law distributions in order to ensure a certain average degree does not appreciably change the procedure of~\cite{CLAUSET:2007} for estimating the exponent.  Suppose $x$ is distributed as a power law lower bound $\xmin$, upper bound $\xmax$, and exponent $\alpha > 0$.  Then
\begin{equation*}
p(x)=\frac{\alpha-1}{\xmin^{1-\alpha}-\xmax^{1-\alpha}}\,x^{-\alpha},\quad \xmin\le x\le \xmax \, .
\end{equation*}
Given a random sample ${\bf x}=\{x_1,\ldots,x_n\}$ drawn from this distribution independently, the likelihood function is
\begin{align*}
p({\bf x})=\prod_{i=1}^n\frac{\alpha-1}{\xmin^{1-\alpha}-\xmax^{1-\alpha}}\,x_i^{-\alpha}
=\left(\frac{\alpha-1}{\xmin^{1-\alpha}-\xmax^{1-\alpha}}\right)^n\prod_{i=1}^nx_i^{-\alpha} \, .
\end{align*}
Thus, the log-likelihood is
\begin{align*}
\log p({\bf x})=n\left(\log(\alpha-1)-\log\left(\xmin^{1-\alpha}-\xmax^{1-\alpha}\right)\right)-\alpha\sum_{i=1}^n\log x_i \, .
\end{align*}
Taking the derivative with respect to $\alpha$ gives
\begin{align}
\label{eq:appc_pdalpha}
\frac{\partial \log p({\bf x})}{\partial \alpha}=n\left(\frac{1}{\alpha-1}+\frac{\xmin^{1-\alpha}\log\xmin-\xmax^{1-\alpha}\log\xmax}{\xmin^{1-\alpha}-\xmax^{1-\alpha}}\right)-\sum_{i=1}^n\log x_i \, .
\end{align}
Setting~\eqref{eq:appc_pdalpha} to zero, we get
\begin{align}
\label{eq:appc_pdalpha_zero}
\frac{1}{\alpha-1}+\frac{\xmin^{1-\alpha}\log\xmin-\xmax^{1-\alpha}\log\xmax}{\xmin^{1-\alpha}-\xmax^{1-\alpha}}=\frac{\sum_{i=1}^n\log x_i}{n} \, .
\end{align}
If $\xmin = 1$ and $\xmax \to \infty$, then solving~\eqref{eq:appc_pdalpha_zero} gives the \MLE\ for $\alpha$ just as in~\eqref{eq:clauset}.


\end{document}